\documentclass[%
 aip, amsmath,amssymb,
jcp, reprint]{revtex4-1}

\usepackage{hyperref}
\hypersetup{
     colorlinks=true,       % false: boxed links; true: colored links
    linkcolor=blue,          % color of internal links (change box color with linkbordercolor)
    citecolor=blue,        % color of links to bibliography
    filecolor=magenta,      % color of file links
    urlcolor=cyan           % color of external links
}

\usepackage[english]{babel}
\usepackage[utf8]{inputenc}
\usepackage[T1]{fontenc}

\usepackage{microtype}

\usepackage{graphicx}

\usepackage{braket}
\usepackage{siunitx}

\newcommand{\rme}{\mathrm{e}}
\newcommand{\rmi}{\mathrm{i}}
\newcommand{\vb}{\boldsymbol}

\newcommand{\rmd}{\mathrm{d}}

\DeclareMathOperator{\diag}{diag}

\begin{document}
\date{\today}

\title{Optimization of selective two-photon absorption in cavity polaritons}
\author{Edoardo G Carnio}
\email{edoardo.carnio@physik.uni-freiburg.de}
\address{Physikalisches Institut, Albert-Ludwigs-Universit\"{a}t Freiburg, Hermann-Herder-Stra{\ss}e 3, D-79104, Freiburg, Germany}
\address{EUCOR Centre for Quantum Science and Quantum Computing, Albert-Ludwigs-Universit\"{a}t Freiburg, Hermann-Herder-Stra{\ss}e 3, D-79104, Freiburg, Germany}

\author{Andreas Buchleitner}
\address{Physikalisches Institut, Albert-Ludwigs-Universit\"{a}t Freiburg, Hermann-Herder-Stra{\ss}e 3, D-79104, Freiburg, Germany}
\address{EUCOR Centre for Quantum Science and Quantum Computing, Albert-Ludwigs-Universit\"{a}t Freiburg, Hermann-Herder-Stra{\ss}e 3, D-79104, Freiburg, Germany}

\author{Frank Schlawin}
\email{frank.schlawin@mpsd.mpg.de}
\address{Max Planck Institute for the Structure and Dynamics of Matter, Luruper Chaussee 149, D-22761 Hamburg, Germany}
\address{The Hamburg Centre for Ultrafast Imaging, Luruper Chaussee 149, D-22761 Hamburg, Germany}
\address{Clarendon Laboratory, University of Oxford, Parks Road, Oxford OX1 3PU, United Kingdom}

\begin{abstract}

We investigate optimal states of photon pairs to excite a target transition in a multilevel quantum system. With the help of coherent control theory for two-photon absorption with quantum light, we infer the maximal population achievable by optimal entangled vs.\ separable states of light.
Interference between excitation pathways, as well as the presence of nearby states, may hamper the selective excitation of a particular target state, but we show that quantum correlations can help to overcome this problem, and enhance the achievable ``selectivity'' between two energy levels, i.e.\ the relative difference in population transferred into each of them. We find that the added value of optimal entangled states of light increases with broadening linewidths of the target states.
\end{abstract}

\maketitle

\section{Introduction}

The theoretical description of (nonlinear) spectroscopy is conventionally based on a semiclassical approach, where the light fields are treated classically and only the sample system is treated fully quantum mechanically~\cite{Mukamelbook, HammZannibook}. In most situations, this approximation is extremely well justified, owing to the weak nonlinearity of light-matter interactions in free space~\cite{Boydbook}. 
Nevertheless, recent years have seen the rapid rise of theoretical investigations, as well as first proof-of-concept experiments that challenge this convention, and investigate how quantum properties of light can be applied or exploited beneficially in spectroscopic applications~\cite{Dorfman2016, Schlawin2018, Gilaberte2019, Mukamel2020, Szoke2020}. This includes the use of photon correlation measurements to analyse the light fields emitted by a sample~\cite{delValle12, GonzalezTudela13, Wientjes14, Holdaway18, Carlos20} in single-molecule spectroscopy~\cite{Kruger10, Hildner13, Kyeyune19}, or to exploit coincidence measurements to detect particular spectral features in the sample~\cite{Dorfman12, Schlawin16, Dorfman18, Zhang18}, as well as the generation of photonic entanglement in fluorescent proteins~\cite{Siyuan17}.

However, the arguably most active field of research concerns the use of quantum light, and in particular of entangled photons, to excite the sample. Squeezed states can improve linear absorption measurements~\cite{Li20,Li20b}. 
Nonlinear optical signals such as two-photon absorption scale linearly with the photon flux~\cite{Georgiades1995, Dayan2004, Lee2006}, which could enable nonlinear spectroscopy of photosensitive samples at very low intensities. Following earlier experiments on two-photon absorption in biomolecules~\cite{Lee2006, Upton2013}, a series of recent experiments have scrutinised the situation~\cite{Parzuchowski2020, Landes2020, Tabakaev2021} and report widely differing entangled two-photon absorption cross sections. They inspired new theoretical investigations into the enhancement that entanglement can provide in two-photon absorption \cite{Raymer2020, Raymer2021}.
Apart from the linear scaling, theoretical proposals show that spectral quantum correlations of entangled photon pairs could further help to disentangle complex optical signals and reveal otherwise hidden features~\cite{Schlawin2013, Raymer2013, Roberto2019}, and enable ultrafast spectroscopy in a cw setup~\cite{Ishizaki2020}. They could also be used for the generation of pseudo-sunlight to imitate natural conditions~\cite{Fujihashi20} and provide sensitive probes for dynamical symmetry breaking~\cite{Li17} and many-body correlations~\cite{Li19}. The control of quantum correlations using temperature~\cite{Roberto2019}, photon statistics~\cite{Castro2019, Csehi19} or spectral shaping~\cite{Oka2018} could further enhance these beneficial properties and provide experimentalists with new handles to manipulate optical signals in a way that is not possible in laser-based spectroscopies.

One pertinent question regarding the application of quantum light to spectroscopy is how much the quantum nature of the former can enhance a given spectroscopic task. 
To ultimately decide this in an unbiased way, it is necessary to compare the performance with optimised (classical) laser pulses, i.e.\ with the optimal performance achievable by classical means. The optimal control with shaped classical laser pulses is a well established field of research~\cite{Rabitz2000, Brif2010}. In particular, optimal control of two-photon absorption was described and implemented experimentally in the late 1990's~\cite{Dubov1996, Assion_1998, Meshulach1998, Meshulach1999, Dudovich2001}.
In these applications, which typically rely on strong laser fields that can manipulate the interference between excitation pathways, quantum light is seen as rather detrimental, as pointed out in Ref.\ \onlinecite{ShapiroBrumer}. However, we recently showed that this is not true for weak broadband fields, where quantum correlations of light can enhance excitation probabilities~\cite{Schlawin2017a, Carnio21}. 

To find these optimal quantum states of light to drive a two-photon transition, we used a coherent control theory for continuous-mode quantum light. In particular, this enabled us to quantify the possible enhancement of the two-photon absorption probability due to quantum correlations between frequency components of the injected pulses, which go hand in hand with strong correlations of the arrival time of the photons.
We have so far restricted this analysis to a simple three-level system. 
In this paper, we extend this theory to two-photon excitations in multilevel systems.

We connect to our previous work by briefly recalling in Sec.\ \ref{sec.theory} the theoretical framework of light-matter interactions. In Sec.\ \ref{sec.optimization}, we recollect the theory of optimal states driving a three-level system, and then generalize this to multilevel targets, where the excitation of undesired, nearby states can prevent that of the particular target states. Finally, in Sec.\ \ref{sec.application} we apply the formalism to a cavity polariton system, where an interesting interference effect involving entangled photons was recently reported in Ref.~\onlinecite{Gu_2020}. We conclude with Sec.\ \ref{sec.conclusions}.

\section{Theoretical framework}
\label{sec.theory}

We consider two pulses -- each carrying a single photon -- impinging on the atomic target. These fields and matter degrees of freedom are described, respectively, by Hamiltonians $H_\text{f}$ and $H_\text{m}$, and are coupled by a light-matter interaction Hamiltonian $W$, which we present in the following paragraphs. The total Hamiltonian thus reads $H=H_\text{f} + H_\text{m} + W$.

Both quantized pulses are injected along fixed and distinct spatial directions, and are described by continuous-mode operators, $a_1 (\omega)$ and $a_2 (\omega)$, respectively\cite{Loudon2000}.
The fields' Hamiltonian, ignoring the vacuum energy, is then $H_\text{f} = \sum_j \int_0^\infty \rmd \omega \, \hbar \omega a_j^\dagger(\omega) a_j(\omega)$.
The positive-frequency electric field operator acting on the Hilbert space of photon $j$ (in the interaction picture with respect to $H_\text{f}$) reads
\begin{equation}
E_j^+(z, t) = \rmi \int_0^\infty \!\! \rmd \omega \; \sqrt{ \frac{\hbar \omega}{4 \pi \epsilon_0 c A} } a_j (\omega) \rme^{\rmi [k (\omega) z - \omega t ]}. \label{eq.field1}
\end{equation}
Here, $z$ is the position along the propagation direction, $A$ the quantization area perpendicular to it, and $c$ the speed of light, and we assume a parallel polarization of the pulses.
We consider the target sample placed at $z = 0$, and much smaller than the wavelength of the light field, such that we can drop the spatial modulation of the field operator \eqref{eq.field1}. 
Furthermore, we only consider field states characterized by narrow pulse shapes, of bandwidth $\Delta \omega$, distributed around a central frequency $\omega_0 \gg \Delta \omega$.
Since all expectation values are calculated with respect to these states,
we can safely extend the range of frequency integration and write the electric field operator \eqref{eq.field1} in the \emph{narrow bandwidth approximation}\cite{Loudon2000} as
\begin{equation}
E_j^+ (t) = \rmi \mathcal{E}_0 \int_{-\infty}^\infty \!\! \rmd \omega \; a_j (\omega) \rme^{- \rmi \omega t},
\end{equation}
where $\mathcal{E}_0 = ( \hbar \omega_0 / 4 \pi \epsilon_0 c A )^{1/2}$ approximates the field normalization of \eqref{eq.field1}. The range of integration $(-\infty,\infty)$ will be assumed from now on.

As for the matter degrees of freedom, we consider a system with a ground state $\ket{g}$, multiple intermediate states $\ket{e_1},\ket{e_2},\ldots$ and multiple final states $\ket{f_1}, \ket{f_2}, \ldots$. With each state $\ket{s}$ we associate an energy $\hbar \omega_s$ (we set $\hbar \omega_g = 0$), an inverse lifetime $\gamma_s$ ($\gamma_g = 0$ since the ground state cannot decay), and a Lorentzian line shape $\mathcal{L}_s(\omega) = (\omega - z_s)^{-1}$, where $z_s = \omega_s - \rmi \gamma_s$. For eigenstates with a finite lifetime, we consider an effective, non-Hermitian\cite{Breuer_2007} matter Hamiltonian $H_\text{m} = \sum_j \hbar z_{e_j} \ket{e_j}\bra{e_j} + \sum_k \hbar z_{f_k} \ket{f_k}\bra{f_k}$.
Adjacent manifolds are dipole-coupled with dipole matrix elements (along the fields' polarization) $\mu_{ge_j}$ and $\mu_{e_j f_k}$, respectively. 

The light-matter interaction Hamiltonian in the rotating wave approximation -- which is certainly valid in the present near-resonant perturbative regime where $\mu \mathcal{E}_0 \ll \hbar \omega_0$ (for any of the dipole matrix elements $\mu$ above) -- and, as the subscript indicates, in the interaction picture with respect to $H_\text{f}+H_\text{m}$, is therefore given by~\cite{Loudon2000}
\begin{equation}
W_I (t) = - V (t) E^{-} (t) - V^{\dagger} (t) E^+ (t), \label{eq.H_int1}
\end{equation}
where the components of the dipole operator which annihilate an electronic excitation may be written as
\begin{align}\label{eq:dip-operator}
 V (t) = & \sum_{j,k} \mu_{ge_j}  \rme^{- \rmi z_{e_j} t} \ket{g}\bra{e_j} \nonumber \\
& + \mu_{e_jf_k}  \rme^{- \rmi (z_{f_k} - z_{e_j}) t} \ket{e_j}\bra{f_k},
\end{align}
and $E^{+ (-)} = E^{+ (-)}_1 + E^{+ (-)}_2$, since we assume that each pulse couples identically to the matter degrees of freedom.

\section{Optimization procedure}
\label{sec.optimization}

\subsection{Single final state}
We seek a two-photon state $\ket{\Phi_f}$ of the incoming fields which optimizes the two-photon transition from the ground state $\ket{g}$ to a single final state $\ket{f}$, via a manifold of $n_e$ intermediate states $\ket{e_j}$.

Mathematically, we consider an initial state $\ket{\Phi_f} \otimes \ket{g}$ in the combined matter-field Hilbert space $\mathcal{H} = \mathcal{H}_\text{f} \otimes \mathcal{H}_\text{m}$ (mirroring the subscripts of the Hamiltonians in the previous section), and determine $\ket{\Phi_f}$ such that the time evolution operator $U_I (t, t_0) $ in the interaction picture of $H_\text{f}+H_\text{m}$ maximises the population of the target state $\ket{0} \otimes \ket {f}$, where both photons have been absorbed in order to drive the matter degrees of freedom to the final state $\ket{f}$. 
As described in detail in Ref.~\onlinecite{Carnio21}, we can find this transition amplitude perturbatively. Tracing out the matter degrees of freedom, the desired transition is mediated by an operator acting on $\mathcal{H}_\text{f}$ alone,
\begin{eqnarray}
\hat{T}_{f} & = & \langle f(t)  \vert U_I (t, t_0) \vert g \rangle \notag \\
& = & \iint T_f(\omega_1, \omega_2) a_1(\omega_1) a_2(\omega_2) \rmd \omega_1 \rmd \omega_2 ,\label{eq:trop}
\end{eqnarray}
with the explicit expression of the matter response function
\begin{align}\label{eq:mrf}
T_{f}(\omega_1, \omega_2) = & \left( \frac{\rmi \mathcal{E}_0}{\hbar} \right)^2 \sum_{j} \mu_{ge_j}\mu_{e_jf} \left[ \mathcal{L}_{e_j}(\omega_1) + \mathcal{L}_{e_j}(\omega_2) \right] \nonumber \\
& \times \mathcal{L}_{f}(\omega_1 + \omega_2) \rme^{- \rmi (\omega_1 + \omega_2)t} .
\end{align}

The matter response function \eqref{eq:mrf} is to be derived for the excitation of the population of $\ket{f}$ at a fixed time $t$, with the interaction $W_I(t)$ turned on at $t_0 \rightarrow -\infty$.
Without loss of generality, we can therefore set $t = 0$. 
The state $\ket{\Phi_f}$ can now be found by variation of the functional~\cite{Schlawin2017a, Carnio21} 
\begin{equation}
J_\mathrm{single} [\ket{\Phi}] = p_{f} - \lambda \left( \braket{\Phi | \Phi} - 1 \right), \label{eq.functional1}
\end{equation}
where $p_f = |\braket{0 | \hat{T}_f | \Phi}|^2$ is the population in the target state $\ket{f}$ created by the absorption of $\ket{\Phi}$, leaving the field in the vacuum state $\ket{0}$. The Lagrange multiplier $\lambda$ constrains the optimization to normalized states\footnote{We remark here that, if we restrict our optimization problem to two-photon states, with one photon per mode, as we do here, the normalization constraint $\braket{\Phi | n_1 n_2 | \Phi}$ used in \onlinecite{Schlawin2017a, Carnio21}, where $n_j$ is the photon number operator for beam $j$, corresponds to $\braket{\Phi | \Phi}$ used here.}. 
Defining the (unnormalized) two-photon state $\ket{T_f} = \hat{T}^\dagger_f \ket{0}$ via \eqref{eq:trop}, the functional \eqref{eq.functional1} is maximized\cite{Schlawin2017a, Carnio21} by the state $\ket{\Phi_f} = \mathcal{N}_f^{-1/2} \ket{T_f}$, whose associated \emph{wave function} reads
\begin{equation}\label{eq:unopt}
\Phi_f(\omega_1, \omega_2) = \braket{0|a_1(\omega_1) a_2(\omega_2) | \Phi_f} = \frac{T_{f}^*(\omega_1, \omega_2)}{\sqrt{\mathcal{N}_{f}}} ,
\end{equation}
given the appropriate normalization $\mathcal{N}_f = \braket{T_f | T_f}$, which we determine analytically in the next subsection.

\subsection{Multiple final states}
We now generalize the formalism to the case of a manifold with $n_f$ target states. Our aim is to find the optimal two-photon state that maximally populates a given final state $\ket{f_1}$, while minimizing the population of all other energetically near-degenerate states $\ket{f_2}, \ket{f_3}, \ldots $, which are equally reachable in terms of the energy of the incoming radiation (and assuming that no selection rule prevents this). To this end, we need to generalize the functional \eqref{eq.functional1} such that the target state's population $p_{f_1}$ is maximized, while the excitation of any other states $\ket{f_{j\neq 1}}$ is penalized: 
\begin{equation}\label{eq:functional_pop}
J[\ket{\Phi}] = p_{f_1} - \sum_{j \neq 1} p_{f_j} - \lambda \left( \braket{\Phi | \Phi} - 1 \right) .
\end{equation}
Notice that enforcing strictly vanishing populations in the states $\ket{f_j} \neq \ket{f_1}$ is prevented by the fact that the matter response functions \eqref{eq:mrf} for different target states $\ket{f_j}$ in general exhibit finite overlap.

In analogy to the previous section, to maximize \eqref{eq:functional_pop} we first define two-photon states
$\ket{T_{f_j}} = \hat{T}_{f_j}^\dagger \ket{0}$.
We remind here that these states are not normalized, but their normalization will be taken care of in \eqref{eq:GEP}. By writing $p_{f_j} = | \braket{T_{f_j} | \Phi}|^2$
the functional \eqref{eq:functional_pop} transforms into
\begin{equation}
J[\ket{\Phi}] = |\braket{T_{f_1} | \Phi}|^2 - \sum_{j \neq 1} |\braket{T_{f_j} | \Phi}|^2 - \lambda \left( \braket{\Phi | \Phi} - 1 \right) .
\end{equation}
The state $\ket{\tilde\Phi}$ that maximizes this functional is found by requiring the variational derivative with respect to the dual state\cite{Carnio21} $\bra{\Phi}$ to vanish:
\begin{equation}
\frac{\delta J}{\delta \bra{\Phi}} = 0 \iff \left( K_{f_1} - \sum_{j \neq 1} K_{f_j} \right)\ket{\Phi} = \lambda \ket{\Phi} ,
\end{equation}
with $K_{f_j} = \ket{T_{f_j}}\bra{T_{f_j}}$. A direct and robust way to solve this eigenvalue problem -- compared, e.g., to the introduction of an orthonormal basis via an orthogonalization procedure --  is to formulate it on the subspace spanned by the non-orthogonal states $\lbrace \ket{T_{f_j}} \rbrace_j$. We then obtain the generalized eigenvalue problem\cite{Bronshtein} (GEP)
\begin{equation}\label{eq:GEP}
\diag (1, -1, -1, \ldots) \cdot \vb v = \lambda \, M \cdot \vb v \quad M_{jk} = \frac{\Sigma_{{f_j} {f_k}}}{\sqrt{\mathcal{N}_{f_j} \mathcal{N}_{f_k}}} ,
\end{equation}
where the matrix $M$ is given by the overlaps
\begin{align}\label{eq:overlaps}
\Sigma_{f_j f_k} & = \braket{T_{f_j} | T_{f_k}} \nonumber \\ 
& = -8 \pi^2 \left( \frac{\mathcal{E}_0}{\hbar} \right)^4 \sum_{m,n} \frac{\mu_{ge_m}\mu_{e_mf_j}\mu_{ge_n}\mu_{e_n f_k}}{( z_{e_m} - z^*_{e_n} 
)( z_f - z^*_{f'}
)} ,
\end{align}
and the normalization constants
\begin{equation}\label{eq:norm}
\mathcal{N}_{f} = \Sigma_{ff} = \frac{8 \pi^2 \mathcal{E}_0^4}{\hbar^4 \gamma_{f}} \sum_{m<n} \frac{\mu_{ge_m}\mu_{e_mf}\mu_{ge_n}\mu_{e_nf}(\gamma_{e_m}+\gamma_{e_n})}{ (\omega_{e_m}-\omega_{e_n})^2 + (\gamma_{e_m}+\gamma_{e_n})^2} .
\end{equation}

We remark here that the original optimization problem \eqref{eq:functional_pop} on the space of square-integrable functions on $\mathbb{R}^2$ has been reduced to an eigenvalue problem of dimension $n_f$ on the target state manifold. All matrices which enter \eqref{eq:GEP} are known analytically, and exclusively depend on the spectral properties (eigenenergies, dipole matrix elements, lifetimes) of the matter degrees of freedom (the factors $\mathcal{E}_0/\hbar$ drop out).

The \emph{selective} optimal two-photon wave function $\tilde \Phi (\omega_1, \omega_2)$ is determined by the eigenvector $\tilde{\vb v}$ associated with the largest eigenvalue of \eqref{eq:GEP}, since we seek to maximize \eqref{eq:functional_pop}. Its components provide the optimal linear combination of the \emph{indistinctive} optimal two-photon wave functions $\Phi_{f_j}$ from \eqref{eq:unopt}:
\begin{equation}\label{eq:GEP_sol}
\tilde \Phi(\omega_1, \omega_2) = \sum_{j} \tilde v_j \frac{T^*_{f_j} (\omega_1, \omega_2)}{\sqrt{\mathcal{N}_{f_j}}} .
\end{equation}
From the eigenvectors we can also immediately compute the maximal final-state populations excited by the selecive optimal state:
\begin{equation}
p_{f_k} = \vert (M \cdot \tilde{\vb v})_k \vert^2 , \label{eq.p_f_k}
\end{equation}
where we remind the reader that the eigenvectors of a GEP are orthonormal with respect to the scalar product induced by the $M$ matrix: $(\vb{v}_j, \vb{v}_k)_M = \vb{v}_j^* \cdot M \cdot \vb{v}_k = \delta_{jk}$.

\section{Application to a near-degenerate manifold}
\label{sec.application}

The coherent superposition \eqref{eq:GEP_sol} of indistinctive optimal states \eqref{eq:unopt} suggests that interference effects between the latter might play an important role in maximizing the target state's population $p_{f_1}$.
However, this information must be encoded in the coefficients $\tilde{v}_j$, whose specific value, even if analytically computable, might not prove insightful.
Therefore, to assess the effectiveness of our optimization method in selectively driving a pre-defined two-photon transition, we want to apply it to a manifold with both near-degenerate and non-degenerate states, as a benchmark.
Such a structure is given, e.g., by the atom-field Hamiltonian describing $N$ atoms ``dressed'' by a quantized cavity mode\cite{Cohen_Tannoudji_1998}. After defining the model, we study how the structure of the optimal two-photon states reflects the different \emph{excitation pathways} $\ket{g} \rightarrow \ket{e_j} \rightarrow \ket{f_k}$, i.e.\ the possible transitions, through the intermediate manifold of $\ket{e_j}$ states, that maximize the yield of the target two-photon transition $\ket{g} \rightarrow \ket{f_k}$. In particular, we discuss the \emph{selectivity} with which a specific state, from a near-degenerate pair, can be excited by entangled or separable states of light.

\begin{table*}[htp]
\begin{center}
\begin{tabular}{|c|c|c|c|c|c|c|c|c|}
\hline
$\omega_r / \omega_0$ & \num{-2.04E-02} & \num{0.698} & \num{0.979} & \num{1.26E+00} & \num{1.58E+00} & \num{1.97E+00} & \num{2.00E+00} & \num{2.37E+00} \\
\hline
$\mu_{rs}$ & $g$ & $e_1$ & $e_2$ & $e_3$ & $f_1$ & $f_2$ & $f_3$ & $f_4$ \\
\hline
$g$ & 0 & \num{1.09E+00} & \num{1.61E-03} & \num{0.939} & 0 & 0 & 0 & 0 \\
$e_1$ & \num{1.09E+00} & 0 & 0 & 0 & \num{0.891} & \num{0.475} & \num{0.705} & \num{0.125} \\
$e_2$ & \num{1.61E-03} & 0 & 0 & 0 & \num{0.757} & \num{-7.95E-02} & \num{-7.34E-02} & \num{0.688} \\
$e_3$ & \num{0.939} & 0 & 0 & 0 & \num{-0.205} & \num{0.515} & \num{0.706} & \num{-0.779} \\
$f_1$ & 0 & 0.891 & 0.757 & \num{-0.205} & 0 & 0 & 0 & 0 \\
$f_2$ & 0 & 0.475 & \num{-7.95E-02} & 0.515 & 0 & 0 & 0 & 0 \\
$f_3$ & 0 & 0.705 & \num{-7.34E-02} & 0.706 & 0 & 0 & 0 & 0 \\
$f_4$ & 0 & 0.125 & 0.688 & \num{-0.779} & 0 & 0 & 0 & 0 \\
\hline
\end{tabular}
\end{center}
\caption{Eigenenergies $\omega_r$ (in units of the cavity mode frequency $\omega_0$, with $\hbar \equiv 1$) of the 
dressed states $\ket{g}$, $\ket{e_j}$, $\ket{f_k}$, $j=1,2,3$, $k=1, \ldots, 4$, of the Hamiltonian \eqref{eq:polaritonic} for $N=2$ two-level systems with eigenfrequencies $\omega_1 = 0.8 \omega_0$ and $\omega_2 = 1.2 \omega_0$, respectively, and their mutual dipole matrix elements $\mu_{rs}$. Due to the weak coupling $g_1,g_2 = \SI{0.14}{\electronvolt} \ll \omega_0 = \SI{1}{\electronvolt} \simeq \omega_1 \simeq \omega_2$, the spectral structure of the uncoupled ($g_1 = g_2 = 0$) dressed system -- one non-degenerate ground state, three degenerate single-excitation states, four degenerate double-excitation states -- remain essentially intact, apart from the degeneracies being lifted.
}
\label{tab:H}
\end{table*}%

\subsection{Dressed-state Hamiltonian}

The $N$-atom-field Hamiltonian we work with reads
\begin{equation}\label{eq:polaritonic}
H = \sum_{n=1}^N \omega_n \sigma_n^\dagger \sigma_n + \omega_0 b^\dagger b + \sum_{n=1}^N g_n (\sigma_n^\dagger + \sigma_n)(b + b^\dagger) ,
\end{equation}
with $\hbar \equiv 1$. 
We consider $N=2$ two-level atoms, each with its individual frequency $\omega_n$ and raising (lowering) Pauli operators $\sigma_n$ ($\sigma^\dagger_n$). Each atom couples with the strength $g_n$ to a monochromatic field, i.e.\ to a quantized harmonic oscillator with frequency $\omega_0$ and annihilation (creation) operator $b$ ($b^\dagger$). When the rotating wave approximation is applied, i.e.\ when $g_n \ll |\omega_0 - \omega_n| \ll \omega_n$, $\forall n$, the counter-rotating terms $\sigma_n b$ and $\sigma_n^\dagger b^\dagger$ can be ignored to obtain the Jaynes-Cummings\cite{Jaynes_1963} (for $N=1$) or Tavis-Cummings\cite{Tavis_1968} ($N>1$) model.
In this work, since $N=2$, we target the two-excitation manifold of the Tavis-Cummings Hamiltonian, which is spanned by the basis states\footnote{The notation indicates the states ($\ket{g}$ or $\ket{e}$) of the two two-level atoms, followed by the number of photons left in the cavity mode.}  $\ket{gg;2}$, $\ket{ge;1}$, $\ket{eg;1}$, and $\ket{ee;0}$, that contain, in total, two shared excitations between field and atomic degrees of freedom.
Diagonalization of the Tavis-Cummings Hamiltonian in this manifold yields the four dressed states $\ket{f_1},\ldots,\ket{f_4}$,
where the two central states $\ket{f_2}$ and $\ket{f_3}$ are degenerate with energy $2\omega_0$; however, when $g_n \sim |\omega_0 - \omega_n|$, the counter-rotating terms of the full Hamiltonian \eqref{eq:polaritonic} lift this degeneracy. The discrimination of either state against the other, then, depends on the competition between their energy separation and their linewidths due to lifetime broadening. When these are similar, we have the ideal scenario to test the ability of our method to selectively excite just one of them.

The dressed electronic states of the atom-field Hamiltonian  are also called (cavity) \emph{polaritons}, since \eqref{eq:polaritonic} constitutes a minimal model of the coupling between photons and the oscillating electric dipoles in a loosely bound crystal \cite{Mahan2000}. In this context, and in the text that follows, states from the single- and double-excitation manifolds are addressed, respectively, as ``polaritons'' and ``bipolaritons''.
Two-photon absorption to the manifold of bipolariton states was recently considered in \onlinecite{Gu_2020}. The transition 
was excited by an entangled biphoton state, created by a cw pump laser, with a frequency sum matching the excitation energy $\omega_1 + \omega_2$ of the targeted two-excitation manifold. It was shown that when the bandwidth of the individual photons, which is determined by the so-called entanglement time, becomes very narrow, such that the biphoton state becomes effectively separable, certain bipolariton states become unaccessible (``dark''\cite{Gu_2020}) due to destructive interference between excitation pathways.

Here, instead, we are interested in the interference between different excitation pathways that manifests in the general spectral structure of the optimal states \eqref{eq:GEP_sol}, with no further constraints beyond those introduced by the functional \eqref{eq:functional_pop}. We therefore gather all spectral information on the (dressed) matter degrees of freedom by diagonalizing the Hamiltonian \eqref{eq:polaritonic} with the maximal number of excitations in the cavity mode fixed at $15$ photons (which is sufficient for numerical convergence). 
To ease comparison, we use the parameter values extracted from Ref.~\onlinecite{Gu_2020}:
$\omega_0 = \SI{1}{eV}$, $\omega_1 = 0.8 \omega_0$, $\omega_2 = 1.2 \omega_0$, $g_1 = g_2 = \SI{0.14}{eV} $.
Of the polaritonic spectrum we consider the non-degenerate ground state $\ket{g}$, the $n_e = 3$ polaritons in the single-excitation manifold, and the $n_f = 4$ bipolaritons in the double-excitation manifold (note that, under the above assumption of sufficiently weak coupling $g_n$, the uncoupled polaritonic and bipolaritonic manifolds remain spectrally well-separated, such that the attributes ``(bi)polaritonic'' still remain meaningful). As in Sec.\ \ref{sec.theory}, we set the origin of our energy scale at $\omega_g$, although, for completeness, we report the unshifted energy levels of the dressed states in Table \ref{tab:H}. To characterize the available excitation pathways from the ground state to the two-excitation manifold, we compute the dipole matrix elements entering \eqref{eq:dip-operator}. 
The element $\mu_{rs} = \sum_n \braket{r | \sigma_n^\dagger + \sigma_n | s}$ is taken along the polarization direction of the electric field, and connects the eigenstates $\ket{r}$ and $\ket{s}$ of \eqref{eq:polaritonic} via the excitation of either two-level system, which we take as aligned dipoles. The values of the dipole matrix elements are also reported in Table \ref{tab:H}.\footnote{Note that, as a peculiar feature of the dressed state scenario, the matrix elements also include transitions via sidebands, since
\begin{equation}
\mu_{rs} = \braket{r | D | s} = \sum_{\alpha \alpha'm} c^{(s)}_{\alpha m} [c^{(r)}_{\alpha' m}]^* \braket{\alpha'; m | D | \alpha; m},
\end{equation}
where $D = \sum_n \sigma^\dagger_n + \sigma_n$ and we expand the dressed states as $\ket{r} = \sum_{\alpha m} c^{(r)}_{\alpha m} \ket{\alpha; m}$, with $\alpha, \alpha' \in \{ gg, ge, eg, ee \}$. For the case considered here of small coupling $g_n$, however, these sideband transitions are orders of magnitude less likely than the resonant transitions we address.}

The last quantity we need to calculate the overlaps in \eqref{eq:overlaps} are the linewidths of the (bi)polariton states. 
As we mentioned above, we wish to assess how well our method can resolve two near-degenerate states in the two-excitation manifold. For this reason, we set, for now, $\gamma_f = 0.01 \omega_0 \approx (\omega_{f_3} - \omega_{f_2})/3$. This implies a significant overlap between the indistinctive optimal pulses \eqref{eq:unopt} that excite each bipolariton individually, as one finds directly with \eqref{eq:overlaps}. For the single-excitation manifold, instead, we take $\gamma_e = \gamma_f/2$, as one would expect for the radiative decay of two uncoupled two-level systems.

\subsection{Optimal states}

\begin{figure*}[t] %  figure placement: here, top, bottom, or page
   \centering
   \includegraphics[width=\textwidth]{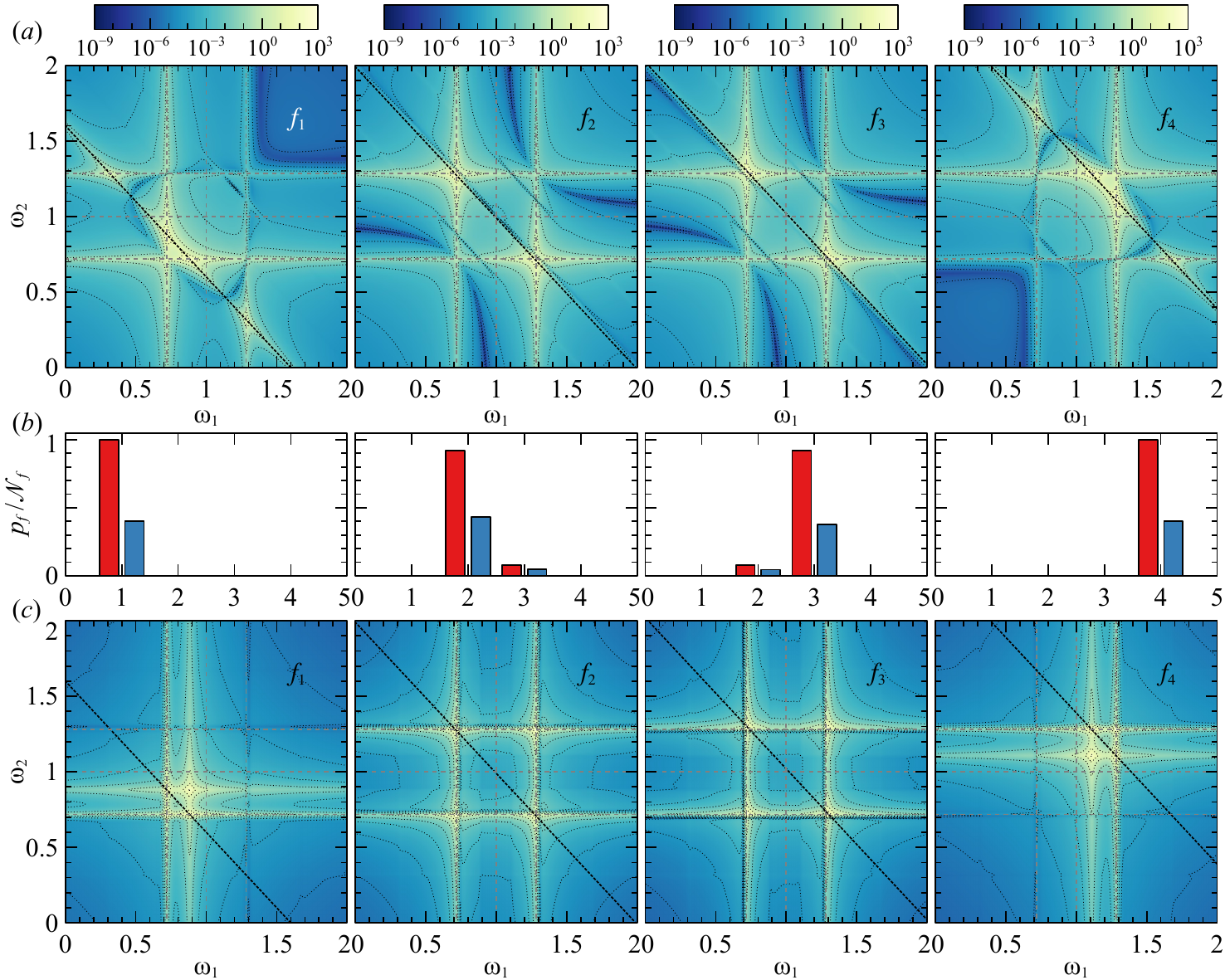} 
   \caption{(\textit{a}) Selective optimal two-photon wave functions \eqref{eq:GEP_sol} exciting, from left to right, the bipolaritons $\ket{f_1}$ to $\ket{f_4}$, as indicated in the top-right corner. Colors and fine-dotted contour lines indicate the intensity of $|\tilde \Phi(\omega_1, \omega_2)|^2$ on a logarithmic scale. Gray dashed lines indicate the resonance frequency of the states $e_j$ in the single-excitation manifold, while black dotted lines delineate the antidiagonal $\omega_1 + \omega_2 = \omega_f$.
   (\textit{b}) Red (left) and blue (right) bars: populations in each bipolariton, in units of $\mathcal{N}_f$ \eqref{eq:norm}, excited by, respectively, the selective optimal state \eqref{eq:GEP_sol} depicted in the panel directly above, and the corresponding classical pulse \eqref{eq:cl_pulse} depicted in the panel directly below. The bars that are not visible are too small compared to the scale, but are never exactly zero.
   (\textit{c}) Classical pulses \eqref{eq:cl_pulse} obtained from the pulses in row (\textit{a}). Colors and fine-dotted contour lines indicate the intensity of $|\tilde \Phi_\text{cl}(\omega_1, \omega_2)|^2$ on a logarithmic scale. Gray dashed lines indicate the resonance frequency of the states $e_j$ in the single-excitation manifold, while black dotted lines delineate the antidiagonal $\omega_1 + \omega_2 = \omega_f$.
   Simulations run with $\gamma_f = 0.01 \omega_0$.
   }
   \label{fig:optimal_wf}
\end{figure*}

The selective optimal two-photon wave functions \eqref{eq:GEP_sol} exciting either one of the bipolariton states $\ket{f_j}$ are distinct. Their moduli square, $| \tilde{\Phi}_{f_j}(\omega_1, \omega_2) |^2$, are all displayed in Fig.\ \ref{fig:optimal_wf}(\textit{a}). 
If the excitation frequency of the target state is $\omega_{f}$, the maxima of the wave functions are aligned along the antidiagonal $\omega_1 + \omega_2 = \omega_{f}$ (black dotted line). Since the excitation of $\ket{e_2}$ is suppressed ($\mu_{ge_2} \approx 0$, see table~\ref{tab:H}), any bipolariton will be reached predominantly by either exciting $\ket{e_1}$ or $\ket{e_3}$ first. 
Hence, we will get maximal population in $\ket{f_1}$, for instance, with photon pairs of frequencies $(\omega_{e_1}, \omega_{f_1}-\omega_{e_1})$ and, to a lesser degree\footnote{Since $\mu_{ge_3}\mu_{e_3 f} < \mu_{ge_1}\mu_{e_1 f}$.}, $(\omega_{e_3}, \omega_{f_1}-\omega_{e_3})$. Consequently, these two points in the two-photon frequency space is where the optimized wave function's density is peaked. 
As discussed at the end of Sec.\ \ref{sec.theory}, the photons couple identically to the matter, which means that either photon can excite the $\ket{g}\rightarrow \ket{e}$ transition, while the other must complete the two-photon absorption by exciting $\ket{e} \rightarrow \ket{f}$. As a consequence, the optimal wave function is symmetric with respect to the two frequencies $\omega_1$ and $\omega_2$. 

The bipolariton population correspondingly excited by each such pulses is given by the red (left) bars in Fig.\ \ref{fig:optimal_wf}(\textit{b}). Note that the populations are given in units of the state normalization $\mathcal{N}_f$ introduced in Eq.~(\ref{eq:norm}). When columns are not visible, the populations are very small compared to the dominating states, but are never exactly zero. From these histograms we see that we can populate almost perfectly the bipolariton states $\ket{f_1}$ and $\ket{f_4}$, which are well separated in energy from competing states. The excitation targeting either $\ket{f_2}$ or $\ket{f_3}$, however, induces a fraction of population also in the respective other state. This is consistent with the two states having a non-negligible overlap for $\gamma_f = 0.01 \omega_0$, as in this regime $\omega_{f_3} - \omega_{f_2} \approx 3 \gamma_f$.

As shown in Refs.~\onlinecite{Schlawin2017a,Carnio21}, the optimal population of the bipolaritons is achieved by the optimized coherent superposition of different frequency modes, such that the atomic response to any frequency pair $(\omega_1, \omega_2)$ adds up constructively. In general, this can only be accomplished if the incoming field modes exhibit entanglement. 
The minimal set of modes required to construct a given entangled state can be computed using the Schmidt decomposition\cite{Nielsen_2009}:
\begin{equation}\label{eq:Schmidt}
\tilde{\Phi}(\omega_1, \omega_2) = \sum_{j=1}^M r_j \phi_j(\omega_1) \psi_j(\omega_2) ,
\end{equation}
which is a weighted sum of $M$ orthonormal modes $\lbrace \phi_j \rbrace_j$ and $\lbrace \psi_j \rbrace_j$. The weights $r_j$ can be chosen real and are listed in decreasing order, by convention. The normalization of the state requires $\sum_{j} r_j^2 = 1$.

The Schmidt modes are useful to construct the optimal \emph{separable} or \emph{classical} state, which contains no quantum correlations:
\begin{equation}\label{eq:cl_pulse}
\tilde{\Phi}_\text{cl}(\omega_1, \omega_2) = \phi_1(\omega_1) \psi_1(\omega_2) .
\end{equation}
It can excite a fraction $r_1^2$ of the optimal population \cite{Schlawin2017a}. This ``classical'' population, which we plot as blue (right) columns in Fig.\ \ref{fig:optimal_wf}(\textit{b}), can be excited in principle by conventionally shaping the frequency spectrum of the individual broadband photons using, for instance, a spatial light modulator to optimize the photonic pulse forms~\cite{Dudovich2001, Dayan2004, Brecht2015}. It cannot however rely on entanglement, i.e.\ on the superposition of different products of modes as in \eqref{eq:Schmidt}.

In Fig.\ \ref{fig:optimal_wf}(\textit{c}) we show the optimal classical states derived from each optimal wave function of row (\textit{a}). The comparison between the two rows clarifies the role of the coherent superposition of the modes in \eqref{eq:Schmidt}: only their constructive or destructive interference can reproduce the profile, dominated by the antidiagonal, of the optimal two-photon wave function. If we consider again the case of $\ket{f_1}$, we see that the classical pulse excites the $(\omega_{e_1}, \omega_{f_1}-\omega_{e_1})$ and $(\omega_{e_3}, \omega_{f_1}-\omega_{e_3})$ transitions, which are resonant with $\ket{f_1}$. 
In addition, however, there is also substantial spectral weight at  $(\omega_{e_1}, \omega_{e_1})$ and $(\omega_{f_1}-\omega_{e_1}, \omega_{f_1}-\omega_{e_1})$, i.e.\ at frequency combinations that are off-resonant for the given $\gamma_f$. In the absence of quantum correlations, they cannot be suppressed. The superposition of Schmidt modes in the entangled state enhances the first two resonant frequency combinations and suppresses the off-resonant combinations, thus enhancing the population by roughly a factor two compared to the classical case [see Fig.\ \ref{fig:optimal_wf}(\emph{b})].

\begin{figure*}[t] %  figure placement: here, top, bottom, or page
   \centering
   \includegraphics[width=\textwidth]{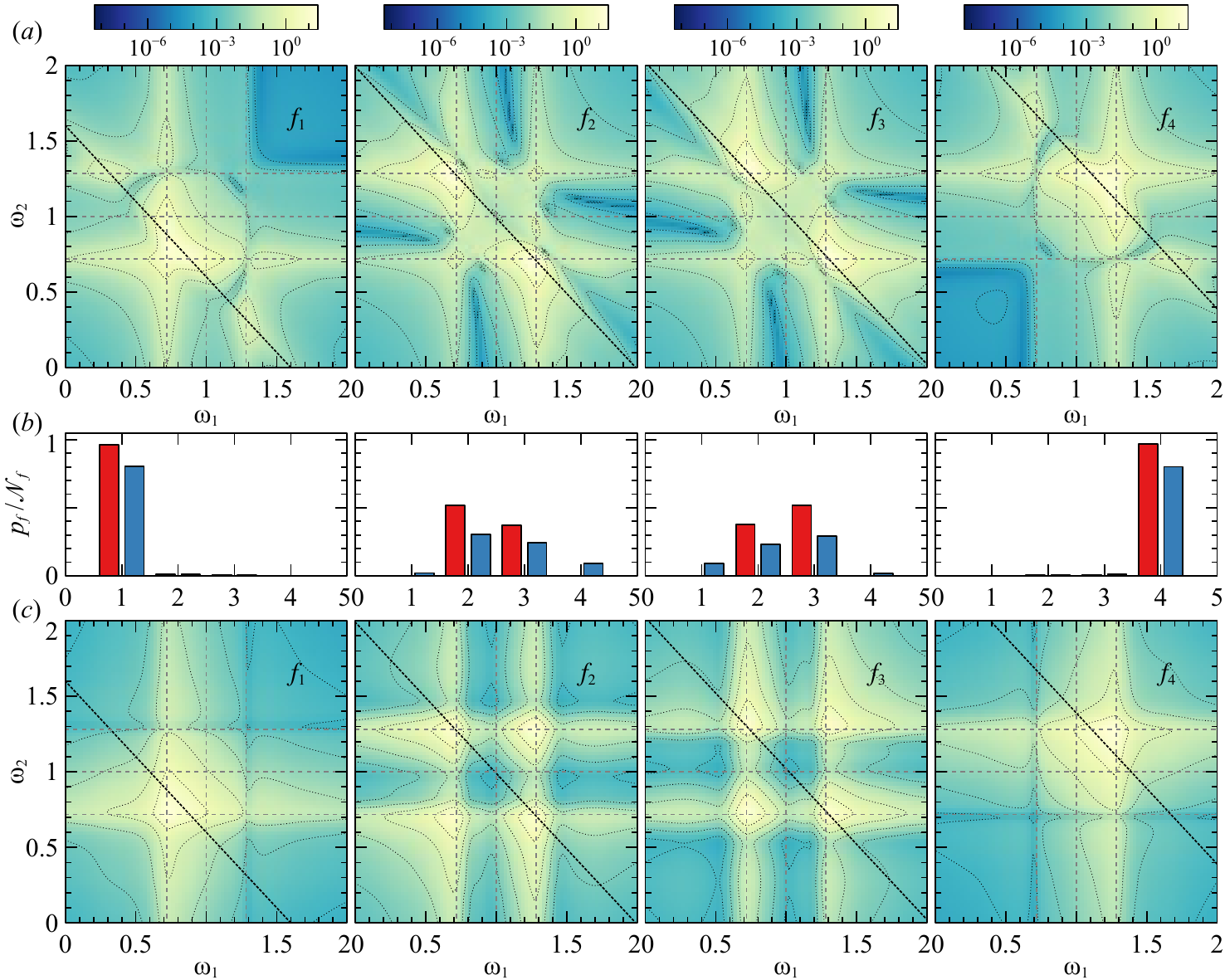} 
   \caption{As in Fig.\ \ref{fig:optimal_wf}: (\textit{a}) Selective optimal two-photon wave functions \eqref{eq:GEP_sol} exciting, from left to right, the bipolaritons $\ket{f_1}$ to $\ket{f_4}$, as indicated in the top-right corner. Colors and fine-dotted contour lines indicate the intensity of $|\tilde \Phi(\omega_1, \omega_2)|^2$ on a logarithmic scale. Gray dashed lines indicate the resonance frequency of the states $e_j$ in the single-excitation manifold, while black dotted lines delineate the antidiagonal $\omega_1 + \omega_2 = \omega_f$.
   (\textit{b}) Red (left) and blue (right) bars: populations in each bipolariton, in units of $\mathcal{N}_f$ \eqref{eq:norm}, excited by, respectively, the selective optimal state \eqref{eq:GEP_sol} depicted in the panel directly above, and the corresponding classical pulse \eqref{eq:cl_pulse} depicted in the panel directly below. The bars that are not visible are too small compared to the scale, but are never exactly zero.
   (\textit{c}) Classical pulses \eqref{eq:cl_pulse} obtained from the pulses in row (\textit{a}). Colors and fine-dotted contour lines indicate the intensity of $|\tilde \Phi_\text{cl}(\omega_1, \omega_2)|^2$ on a logarithmic scale. Gray dashed lines indicate the resonance frequency of the states $e_j$ in the single-excitation manifold, while black dotted lines delineate the antidiagonal $\omega_1 + \omega_2 = \omega_f$.
   Simulations run with $\gamma_f = 0.1 \omega_0$.}
   \label{fig:optimal_wf_0.1}
\end{figure*}

Since the natural linewidth $\gamma_f$ of the bipolaritons determines how close to resonance the two-photon transitions are, Fig.\ \ref{fig:optimal_wf_0.1} shows the optimal two-photon wave functions, the target state population histograms and the accordingly shaped classical pulses for bipolaritons with a linewidth enlarged by a factor ten (i.e.\ shorter lifetimes), $\gamma'_f = 0.1 \omega_0 = 10 \gamma_f$. As one would expect intuitively, while the general structure of the optimal states can still be recognized, given our knowledge of their structures for narrow linewidths as depicted in Fig.\ \ref{fig:optimal_wf}, some details are washed out by the broadened resonances. For the state $\ket{f_1}$, for instance, the square-like structure, visible in Fig.\ \ref{fig:optimal_wf}, of the classical pulse, for $\gamma_f$, cannot be resolved anymore for $\gamma'_f$, because the single-photon frequency distributions are much broader than the difference $(\omega_{f_1} - \omega_{e_1}) - \omega_{e_1} = \omega_{f_1} - 2\omega_{e_1}$. This explains why the population induced by the shaped classical state is larger for $\gamma'_f$ than for $\gamma_f$. Similarly, when we try to optimally populate $\ket{f_2}$, we now obtain also a larger population in $\ket{f_3}$, as well as some in $\ket{f_4}$. The latter is due to the classical pulse having a significant peak around $(\omega_{e_3}, \omega_{e_3})$, which is broad enough to excite the $(\omega_{e_3},\omega_{f_4}-\omega_{e_3})$ two-photon transition to $\ket{f_4}$.

\subsection{Selectivity}
The previous subsection showed that, if the linewidth of the states is much larger than the frequency differences involved in the two-photon transitions, (i) the quantum advantage due to the frequency entanglement in the optimal pulses is reduced, (ii) the achievable populations in $\ket{f_1}$ and $\ket{f_4}$ decrease (increase) with increasing $\gamma_f$ when using an entangled (classical) two-photon state, and (iii) the target population in $\ket{f_2}$ ($\ket{f_3}$) decreases with increasing $\gamma_f$, while that of $\ket{f_3}$ ($\ket{f_2}$) increases. This latter point, in particular, implies that the \emph{selectivity} of the optimal pulse \eqref{eq:GEP_sol}, captured by the contrast between the populations of $\ket{f_2}$ and $\ket{f_3}$,
\begin{equation}\label{eq:contrast}
S = \frac{|p_{f_2}-p_{f_3}|}{p_{f_2}+p_{f_3}} ,
\end{equation}
worsens for broader linewidths.

\begin{figure} %  figure placement: here, top, bottom, or page
   \centering
   \includegraphics[width=\columnwidth]{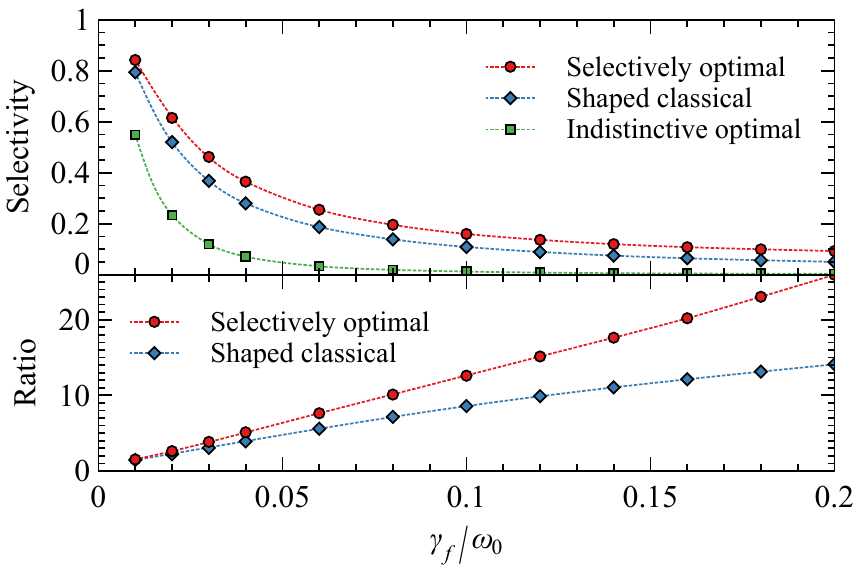} 
   \caption{Top: selectivity $S$ from Eq.\ \eqref{eq:contrast} calculated for selectively optimal [\eqref{eq:GEP_sol}, red circles], shaped classical [\eqref{eq:cl_pulse}, blue diamonds] and indistinctive optimal pulses [\eqref{eq:unopt}, green squares]. Bottom: ratio of the selectivity achieved with selectively optimal (red circles) or shaped classical pulses (blue diamonds), to the selectivity achievable with an indistinctive optimal pulse.
   }
   \label{fig:contrast}
\end{figure}

Finally, an analysis of how the desired selectivity scales with the intermediate and target states' linewidths ($\gamma_f = 2 \gamma_e$), for optimal quantum vs classical two-photon states, reveals the advantage of performing the optimization presented in this work. In Fig.\ \ref{fig:contrast} we compare the selectivity in those cases where the population of $\ket{f_2}$ is induced by selective optimal entangled states \eqref{eq:GEP_sol} to that achieved by shaped classical states of light \eqref{eq:cl_pulse}. In addition, we compare these results to those achieved with the indistinctive optimal quantum state \eqref{eq:unopt}, which, we remind from Sec.\ \ref{sec.optimization}, is optimized to excite the $\ket{g} \rightarrow \ket{f_2}$ transition most efficiently, without the additional constraint to minimize the excitation of other nearby bipolaritons from the same manifold. 
For all three possible injected two-photon states, the selectivity decays with $\gamma_f$, in a qualitatively exponential fashion (top panel). Yet, \emph{selective} optimization as conceived in Sec.\ \ref{sec.optimization}, for classical as well as for quantum light, slows this decay down. For instance, at $\gamma_f = 0.05 \omega_0$, the indistinctive optimal (quantum) pulse will hardly discriminate $\ket{f_2}$ against $\ket{f_3}$ any more, while the optimization procedure can still achieve ca.\ $25$~\% selectivity. 
This relative improvement is visualized by plotting the ratio of selective vs.\ indistinctive yields in the figure's bottom panel: this ratio actually \textit{increases} with increasing $\gamma_f$. This has a very intuitive explanation: the possibility of constructing suitable linear superpositions \eqref{eq:GEP_sol} of indistinctive optimal states \eqref{eq:unopt} relies, in the first place, on their overlap, which increases with $\gamma_f$.

\section{Conclusions}
\label{sec.conclusions}
We have investigated continuous-mode two-photon states to populate a target matter state reachable by two-photon absorption. 
The method discussed in this work optimizes the excitation's selectivity, i.e.\ it maximizes the target state population while minimizing residual population within the complement of the target state's manifold.
The optimal quantum state of light to drive this transition can be obtained by solving a generalized eigenvalue problem where all matrices depend analytically on the spectral properties of the driven system, and have a dimension given by that of the manifold the target state is embedded in.

We have applied our method to the specific setting of two non-interacting atoms dressed by a cavity mode, with the goal of driving the transition from the ground state to one of the four bipolariton states in the two-excitation manifold. If the bipolariton states are well separated in energy, our procedure is equivalent to the optimization of a single pathway analyzed in previous work\cite{Schlawin2017a, Carnio21}. If they overlap, instead, the excitation of an individual transition inevitably induces transitions to the nearby states, too.
In this case, however, we manage to obtain an appreciable selectivity even when it would be impossible to otherwise discern closely neighboring resonances.
The Hamiltonian~(\ref{eq:polaritonic}) considered in this study is formally similar to excitonic models of molecular aggregates\cite{May_2011} in physical chemistry, in the sense that, in the parameter regime considered here, we obtain energetically well-separated manifolds of states, where adjacent ones are dipole-coupled.
It seems highly likely that the excitation physics presented here will carry over to entangled two-photon excitation of, e.g., molecular aggregates~\cite{Schlawin2013, Raymer2013}. 
We note, however, that the inevitable coupling of the electronic states to environmental degrees of freedom will induce additional relaxation processes such as incoherent electronic population transfer.
These processes are not captured by the non-Hermitian Hamiltonian considered here, and would rather require
an open systems description, where the material degrees of freedom evolve according to a master equation. 

While the solution of the selective optimization problem is general, its benchmark inevitably depends on the specific structure of the target spectrum considered. In particular, to enable a comparison with \onlinecite{Gu_2020}, we considered a sufficiently small coupling constant $g_n$ in \eqref{eq:polaritonic}, such that the single- and double-excitation manifolds remain well separated in energy. The selective excitation of transitions in a spectrum where said manifolds mix, instead, will be a topic of future research.

\begin{acknowledgements}
This work is supported by the European Research Council under the European Union's Seventh Framework Programme (FP7/2007-2013) Grant Agreement No.~319286 Q-MAC. 
E.G.C.~acknowledges support from the Georg H.~Endress foundation.
F.S.~acknowledges support from the Cluster of Excellence `Advanced Imaging of Matter' of the Deutsche Forschungsgemeinschaft (DFG) - EXC 2056 - project ID 390715994.
\end{acknowledgements}

\section*{DATA AVAILABILITY}
The data that supports the findings of this study are available within the article.

\bibliography{jcp_refs.bib}

\end{document}